\documentclass[a4paper,onecolumn,12pt]{article}
\usepackage{amsmath}
\usepackage{amsfonts}
\usepackage[bookmarks=true, bookmarksnumbered=true, linkbordercolor={1 1 1}]{hyperref}
\usepackage{float}
\usepackage{placeins}
\usepackage[cal=boondoxo]{mathalfa}
\usepackage{tikz}
\usetikzlibrary{fit,positioning}
\usepackage{lscape}
\usepackage{longtable}
\usepackage{epstopdf}
\usepackage{float}
\usepackage{setspace}
\usepackage[mathscr]{euscript}

\usepackage{algorithm}
\usepackage[end]{algpseudocode}
\algnewcommand\algorithmicto{\textbf{to}}
\usepackage{etoolbox}
\patchcmd{\thebibliography}{\section*{\refname}}{}{}{}

\usepackage{titlesec}
\titleformat{\section}
{\normalfont\normalsize\bfseries}{\thesection.}{1em}{}
\titleformat{\subsection}
{\normalfont\normalsize\bfseries\itshape}{\thesubsection.}{1em}{}
\titleformat{\subsubsection}
{\normalfont\normalsize\itshape}{\thesubsubsection.}{1em}{}
\titleformat{\subsubsubsection}
{\normalfont\normalsize\itshape\underline}{\thesubsubsection.}{1em}{}

\begin{document}

\title{\LARGE{Machine Learning Econometrics: Bayesian algorithms and methods}}
\author{\textbf{Dimitris Korobilis} \\
University of Glasgow  \and \textbf{Davide Pettenuzzo} \\
Brandeis University}

\date{\today}

\maketitle
\begin{abstract}
\noindent As the amount of economic and other data generated worldwide increases vastly, a challenge for future generations of econometricians will be to master efficient algorithms for inference in empirical models with large information sets. This Chapter provides a review of popular estimation algorithms for Bayesian inference in econometrics and surveys alternative algorithms developed in machine learning and computing science that allow for efficient computation in high-dimensional settings. The focus is on scalability and parallelizability of each algorithm, as well as their ability to be adopted in various empirical settings in economics and finance.

 \end{abstract}
\bigskip

\noindent \textbf{JEL Classification:} C11; C15; C49; C88 \\
\\
\noindent \textbf{Keywords:} MCMC; approximate inference; scalability; parallel computation

\newpage

\tableofcontents
\newpage

\onehalfspacing
\section{Introduction and background}

The purpose of this review is two-fold. The first aim is for this to be an accessible reference of various algorithms that economists can use for inference problems in the Big Data era. The second aim is to introduce methods and algorithms developed outside economics (e.g. computing science, machine learning, engineering) and discuss how economists can benefit from this wealth of work done by other scientists. The primary focus is on Bayesian algorithms, even though in many cases the algorithms analyzed are appropriate for maximum likelihood inference. Bayesian methods have been traditionally used in econometric problems that either involve complex likelihood structures or a large number of variables relative to observations. Such examples are the class of dynamic stochastic general equilibrium (DSGE) models, panel data with fixed effects or cross-sectional data with many predictors (e.g. growth regressions). In particular, Monte Carlo methods have allowed to simplify even the toughest of inference problems. However, existing Monte Carlo techniques such as the Gibbs sampler or Metropolis-Hastings algorithms are inherently demanding and can quickly hit a computational bottleneck. Therefore, a major question that this review attempts to answer is the following: what other options are there for speeding up Bayesian inference when faced with high-dimensional models and data?

The starting point for Bayesian estimation and computation is Bayes rule
\begin{equation}
p(\theta | y ) = \frac{p(y | \theta) p (\theta)}{\int p(y | \theta) p (\theta) d\theta},
\end{equation}
where $\theta$ represents the parameters of our chosen model we want to estimate, $p(y | \theta)$ is the likelihood function of the specified model, $p (\theta)$ is the function of parameters before seeing the data (prior), and $p(\theta | y )$ the distribution of the parameters after observing the data (posterior). The quantity $\int p(y | \theta) p (\theta) d\theta$ is called the marginal likelihood and is a constant that ensures that the posterior has a density that integrates to one.\footnote{When one wants to calculate the posterior analytically this integral needs to be evaluated numerically. Otherwise when sampling methods are used we might only need to know the kernel of the posterior, in which case we simply use the expression $p(\theta | y ) \propto p(y | \theta) p (\theta)$.} The idea here is that parameters are random variables, despite the fact that Bayesian consistency requires that in the limit (infinite observations) $\theta$ should converge to the true point parameter $\theta_0$. 

Maximum likelihood (ML) inference would require us to work only with $ p(y | \theta)$, however, maximizing complex functions (e.g. a high-dimensional, nonlinear likelihood) is not a computationally trivial task. Instead, as Angelino et al. (2016) observe, the Bayesian paradigm is about integration. The Bayesian needs integration in order to compute marginal and conditional posteriors, prior predictive distributions (marginal likelihoods) for model comparison and averaging, and posterior predictive distributions for making predictions. 

Needless to stress that in high dimensions integration doesn't become computationally more desirable than maximization used in the ML approach! So what are the relevant Bayesian tools that a modern economist could and should have in her toolbox in order to perform Bayesian inference in high-dimensions? Some key estimation algorithms that econometricians and economists have been using already for decades, are reviewed in the next Section. Subsequently, Section 3 covers several algorithms developed in fields such as computer vision, signal processing, and compressive sensing, among other fields that rely on analysis of high-dimensional data. Finally, recommendations are provided on specific ways of speeding up Bayesian inference by simplifying an econometric model in such a way that one can get ``more mileage'' from Bayesian algorithms.

\section{A review of Bayesian computation}

\subsection{Exact and approximate analytical results}
\subsubsection{Uniform and conjugate priors}
There are only a handful of cases of prior distributions that, when multiplied by a likelihood function, allow for analytical derivation of the posterior distribution and all its moments. In standard linear regression settings, uniform and natural conjugate priors allow for working with posterior distributions that belong to well known classes (Normal, Gamma, Wishart). The uniform prior collapses to multiplying the likelihood by a constant, such that the posterior is proportional to the likelihood, and the posterior mode becomes identical to the maximum likelihood estimate. The natural conjugate prior for regression models with coefficients $\beta$ and variance parameter $\sigma^{2}$ has the form
\begin{equation}
p \left (\theta \right) \equiv p \left( \beta, \sigma^2 \right) = p(\beta \vert \sigma^{2}) p(\sigma^{2}).
\end{equation}
The ``unnatural'' feature of the natural conjugate prior formulation is that we need specify our prior for $\sigma$ independently, but our prior opinion about $\beta$ is conditional on the values of $\sigma$. Nevertheless, such priors lead to an analytical expression for the parameter posteriors, that is, posterior means, variances, and other moments are readily available in closed form. This is the reason why such priors were widespread many decades ago, well before cheap and strong computing became available. Interestingly, recent econometric papers have revived interest in using such simple priors, by exploiting their simplicity in order to estimate effortlessly large vector autoregressions (VARs) with hundreds of thousands of coefficients; see the discussion in Korobilis and Pettenuzzo (2019).

\subsubsection{Normal and Laplace approximations}
In more complex settings where conjugate priors cannot be defined, the posterior can sometimes be approximated by a Normal distribution. According to the Bayesian central limit theorem, under certain conditions, the posterior distribution $p \left( \theta \vert y \right)$ is asymptotically Normal. The Bernstein-von Mises theorem states that the posterior distribution is asymptotically independent of the prior distribution, thus, giving further justification to a Normal approximation of the posterior distribution.

Laplace (1774) was the first to argue that for any continuous posterior that is smooth and well-peaked around its point of maxima (mode), a Normal approximation is a sensible choice. First, note that if $\theta^{\star} = \arg \displaystyle \max_{\theta \in \Theta} p \left(\theta \vert y \right)$ is the maximum of the posterior function, then this will also be the maximum of the log-posterior $h\left( \theta \right) = \log \left( p \left( \theta \vert y \right) \right)$. Then a second-order Taylor series expansion of the log-posterior around $\theta^{\star}$ gives
\begin{align}
\begin{split}
h\left( \theta \right) & \approx  h\left( \theta^{\star} \right) + \dot{h} \left( \theta^{\star} \right) \left( \theta - \theta^{\star} \right) - \frac{1}{2} \left( \theta - \theta^{\star} \right)^{\prime} \ddot{h} \left( \theta^{\star} \right) \left( \theta - \theta^{\star} \right), \\
& \approx   const - \frac{1}{2} \left( \theta - \theta^{\star} \right)^{\prime} \ddot{h} \left( \theta^{\star} \right) \left( \theta - \theta^{\star} \right),
\label{eq_taylor_approx}
\end{split}
\end{align}
where $\dot{h} \left( \theta^{\star} \right)$ and $\ddot{h} \left( \theta^{\star} \right)$ are the first and second derivatives of the log-posterior function. Given that $\theta^{\star}$ is a maximum, it follows that $\dot{h} \left( \theta^{\star} \right) = 0$, which justifies the simplification in the second row of equation \eqref{eq_taylor_approx}. Similarly, $\ddot{h} \left( \theta^{\star} \right)$ is positive definite, which implies that the log-posterior is proportional to a Normal kernel. Equivalently, by taking the exponential function on both sides of the expression in equation \eqref{eq_taylor_approx} we have
\begin{equation}
p \left( \theta \vert y \right) = \exp \left( h \left( \theta \right) \right) \sim N \left( \theta^{\star},  \ddot{h} \left( \theta^{\star} \right) \right),
\end{equation}
which provides a justification for a Normal approximation to the posterior. Therefore, instead of integrating to find the posterior, the Bayesian inference problem becomes an optimization one: once we find $\theta^{\star}$ and $\ddot{h} \left( \theta^{\star} \right)$, we have everything we need in order to describe the (approximate) posterior analytically.\footnote{As with maximum likelihood or maximum a-posteriori (MAP) inference (see \autoref{MAP}) one can use a range of well-known numerical optimization routines to find the posterior mode $\theta^{\star}$ (e.g. quasi-Newton methods).} The approximation error of the Laplace approach is $O\left( N^{-1/2}
 \right)$. If the posterior is asymmetric or skewed, then including higher-order derivatives of $h \left( \theta \right)$ in the Taylor series expansion can improve the approximation (Lindley, 1980). However, evaluating numerically such terms for the log-posterior function can be impractical computationally. Laplace approximations are fast, accurate and easy to implement. Nevertheless, in high-dimensional problems it becomes difficult, if not impossible, to numerically evaluate the joint posterior mode because the posterior function could be too complex and multi-modal. 

\subsubsection{Bayesian Quadrature and Monte Carlo}
Assuming that the parameter vector $\theta$ has $K$ elements, Naylor and Smith (1982) note that the marginal posterior of $\theta_{i}$, $i=1,...,K$, is of the form
\begin{equation}
p \left( \theta_{i} \vert y \right) = \int p(y | \theta) p (\theta) d \theta_{j \neq i}, \label{marginal}
\end{equation}
where $d \theta_{j \neq i}$ denotes integration over the $K-1$ terms $\theta_j$, $j=1,...,K$ for $j \neq i$. As discussed later in this review, many modern Bayesian machine learning algorithms exploit this result and work with the marginal posterior distribution. This is because the $K$ marginals $p \left( \theta_{i} \vert y\right)$ can be trivially processed in parallel using modern multi-core systems. Of course, this was not the initial intention of the early work of Naylor and Smith (1982). Rather their focus on the marginal posterior in equation \eqref{marginal} was driven by their desire to use iterative quadrature methods for estimating such integral. Naylor and Smith (1982) in particular suggest an adaptive Gauss-Hermite quadrature, while others have proposed Gaussian process (GP) priors\footnote{GP priors are priors over functions and their values.} leading to the ``Bayesian quadrature'' algorithm. Alternatively, the integral in equation \eqref{marginal} can be evaluated using Monte Carlo Integration. Rasmussen and Ghahramani (2003) argue that classical Monte Carlo estimators violate the \emph{Likelihood Principle} and instead propose a Bayesian Monte Carlo procedure.

\subsection{Importance sampling}
A natural question is what should a Bayesian do if she derives an expression for the posterior distribution that is not in a form that she recognizes or can easily be sampled from (e.g. Normal, Bernoulli, Gamma or any other distribution that we can sample from easily). Under this scenario, importance sampling offers a very intuitive and simple solution: if you do not recognize $p(\theta \vert y)$, choose instead a ``proposal distribution'' $q(\theta)$ that is easy to sample from and convert its samples into samples from the desired density $p(\theta \vert y)$. Assume we collect $n$ such draws, $\widehat{\theta}^{(1)},...,\widehat{\theta}^{(n)} \sim q$. Next, estimate weights $w^{(i)} = \frac{p(\widehat{\theta}^{(i)} \vert y)}{q(\widehat{\theta}^{(i)})}$,\footnote{It is important to note that, henceforth, $p(x)$ denotes a distribution function for random variable $x$, and $p(\widehat{x})$ denotes the same distribution $p$ evaluated at the value $\widehat{x}$. The latter is going to be a number (probability), and the difference between the two expressions stems from the fact that once values $\widehat{x}$ are sampled (observed), these are not random variables any more.} for $i = 1,...,n$, and use them to obtain the importance weighted estimator
\begin{equation}
\widetilde{\theta}  = \frac{\sum_{i=1}^{n} w^{(i)} \widehat{\theta}^{(i)}}{\sum_{i=1}^{n} w^{(i)}}.
\end{equation}
As long as the support of $q$ contains the support of $p(\theta \vert y)$, it can be shown that $\widetilde{\theta}$ converges to $E \left(\theta \vert y \right)$; see Geweke (1989) for detailed results. Unfortunately, when $\theta$ is high-dimensional it can be very hard to find a $q$ that meets this condition, hence importance sampling becomes harder to implement in very large models.

\subsection{Metropolis-Hastings algorithm}
Metropolis-Hastings is a class of Monte Carlo algorithms based on accept/reject sampling, that extends ideas in importance sampling. Assume we have obtained $S$ samples from a proposal distribution $q$ and further assume that the $(i-1)^{th}$ sample we generated, denoted by $\widehat{\theta}^{(i-1)}$, is indeed a sample from $p(\theta \vert y)$. Finally, denote with $\widehat{\theta}^{\star}$ the $i$-th candidate sample from $q$. Then $\widehat{\theta}^{\star}$ is accepted with probability
\begin{equation}
\alpha \left( \widehat{\theta}^{\star},\widehat{\theta}^{(i-1)} \right) = min \left\lbrace 1, \frac{p(\widehat{\theta}^{\star} \vert y)q(\widehat{\theta}^{(i-1}) \vert \widehat{\theta}^{\star})}{p(\widehat{\theta}^{(i-1)} \vert y)q(\widehat{\theta}^{\star} \vert \widehat{\theta}^{(i-1)})}.  \right\rbrace \label{MH_prob}
\end{equation}
If the acceptance ratio $\alpha$ is larger than  a random draw $u$ from a $Uniform(0,1)$ then we accept the draw and set $\widehat{\theta}^{(i)} = \theta^{\star}$, otherwise we discard it and set $\widehat{\theta}^{(i)} = \widehat{\theta}^{(i-1)}$.

In order to guarantee that draws $\widehat{\theta}^{(i)}$ are samples from the target posterior $p(\theta \vert y)$ we aim to approximate, we need several desirable features for this chain such as irreducibility and aperiodicity. Chib and Greenberg (1995) offers an early, accessible reference to Metropolis-Hastings. Applications of the MH algorithm are numerous in economics, with most notably its use in nonlinear state-space formulations for the purpose of estimating dynamic stochastic general equilibrium (DSGE) models. As with importance sampling, the Metropolis-Hastings algorithm can become inefficient in very large dimensions, with low rates of acceptance, poor mixing of the chain and highly correlated draws.

\subsection{Gibbs sampler}
With the Gibbs sampler the aim is to sample from the conditional posterior, that is, the posterior of each parameter conditional on all other model parameters being fixed to a known value. Assume that $\theta$ has $n$ elements or blocks, $\theta_{1},...,\theta_{n}$, e.g. in the most plain univariate regression with one regressor this would be $\theta_{1} = \beta$ and $\theta_{2} = \sigma^{2}$. Thanks to a straightforward application of Bayes Theorem, it holds that samples from the conditional posteriors are also samples from the joint parameter posterior
\begin{equation}
p(\theta_{j} \vert \theta_{1},...,\theta_{j-1},\theta_{j+1},...,\theta_{n},y) = \frac{p( \theta_{1},...,\theta_{n} \vert y)}{p(\theta_{1},...,\theta_{j-1},\theta_{j+1},...,\theta_{n} \vert y)} \propto p( \theta_{1},...,\theta_{n} \vert y) \equiv p(\theta \vert y).
\end{equation}
The conditional posterior for each $\theta_j$ is proportional to the joint posterior simply because the denominator is a constant (all $\theta_{k}$ for $k \neq j$ are conditioned upon and are known/fixed, hence, $p(\theta_{1},...,\theta_{j-1},\theta_{j+1},...,\theta_{n} \vert y)$ is the value of the p.d.f.). The Gibbs sampler can be viewed as a special case of Metropolis-Hastings algorithms where every draw is accepted with probability one: if we assume that the conditional posterior $p(\theta_{j} \vert \theta_{1},...,\theta_{j-1},\theta_{j+1},...,\theta_{n},y)$ $\forall \text{ \ } j$ is the proposal density $q$, then it is trivial to show via equation \eqref{MH_prob} that $\alpha = min \left\lbrace 1,1 \right\rbrace$.

The Gibbs sampler is probably the most user-friendly among the class of MCMC algorithms. It simplifies computation of some complex econometric and statistical models that would otherwise be extremely hard to estimate with maximum likelihood. Deriving a conditional posterior involves an expression for a parameter $\theta_{j}$ by keeping other parameters fixed (to their last sampled values), an idea that is most useful in nonlinear and latent parameter models. For instance, consider the example a Markov switching autoregression (AR) for measuring business cycles: conditional on knowing the indicator variables indexing the Markov states, Gibbs sampler inference on the autoregressive coefficients and the variance parameter is identical to that of the standard AR model. Furthermore, more complex nonlinear problems can be easily transformed to linear, Gaussian problems that can be approximated trivially by the Gibbs sampler; see, among others, the well-known estimator for stochastic volatility models of Kim et al. (1998).

\newpage
\section{Bayesian methods in the Big Data Era}

As we adjust to the new reality of having larger amounts of data available, the Bayesian computation methods that we have briefly reviewed in the previous section also need to be adapted and improved. In particular, as the data size, number of features, size of the models, and model space all growth, it becomes computationally harder to evaluate the likelihood function, visiting all the parameters in the model, while at the same time using all the data. At the same time, the algorithms that we discussed previously also start to experience slower mixing rates. One may therefore be sceptical about the possibility of adapting Bayesian methods to keep up with this trend. However, it is worth noting that Bayesian methods features a number of important advantages that make them particularly appealing even in the Big Data Era. First and foremost, Bayesian methods offer the flexibility and adaptivity required to deal with a reality in which the volume of the data grows. The  updating rule which is at the core of Bayesian methods is sequential in nature and suitable for dealing with constantly growing data streams. 

\medskip 

The main complication of applying Bayesian methods to big data problems has to do with the computational bottlenecks that the previously described algorithms face, and for that reason existing literature has been hard at work developing new (approximate) methods to deal with this evolving reality. In this Section, these issues are discussed in more detail and some of the solutions that have been proposed in the literature to deal with the increasing amounts of data and the larger computational costs that researchers face when implementing Bayesian methods, are reviewed.

\subsection{Speeding up MCMC}
The first step towards making Bayesian computation feasible in a high-dimensional setting, is to use approximations that replace computationally intensive steps of MCMC algorithms. One solution proposed in the literature is to use approximate samplers that use data sub-samples (\emph{minibatches}) rather than the full data set. Examples include subsampling Markov chain Monte Carlo (MCMC) implementations; see Bardenet, Doucet, and Holmes (2017) for an excellent review of these approaches. The basic idea is to to estimate the likelihood function for $n$ observations from a random subset of $m$ observations, where $m \ll n$. With conditionally independent observations, one can rewrite the log-likelihood $\mathcal{l} \left(\theta \right)=\log p\left(\left. y \right\vert \theta \right) $ as follows
\begin{align}
\mathcal{l}	\left(\theta \right) = \sum_{i=1}^n \mathcal{l}_i	\left(\theta \right) \label{loglik_decomp}
\end{align}
where $\mathcal{l}_i	\left(\theta \right)=\log p\left(\left. y_i \right\vert \theta \right)$ denotes the log-likelihood contribution of the $i$-th observation in the sample.\footnote{The assumption that the total log-likelihood can be decomposed into a sum of terms where each term depends on a unique piece of information is not overly restrictive. It applies to longitudinal problems but also to certain time series problems such as AR($p$) processes.} As it turns out, estimating \eqref{loglik_decomp} using simple random sampling where any $\mathcal{l}_i	\left(\theta \right)$ is included with the same probability, generally results in a very large variance. This problem could be eliminated if one were to re-weight the draws using so called probability proportional-to-size sampling, but unfortunately computing these weights can be computationally very expensive. One way to sidestep this computational bottleneck is to make the $\left\{\mathcal{l}_i	\left(\theta \right)\right\}_{i=1}^n$ more homogeneous by using control variates so that the population elements are roughly of the same size. In this way, a simple random sampling would then expected to be efficient. This is the approach taken by Quiroz et al (2019), who  use control variates to obtain  a highly efficient unbiased estimator of the log-likelihood, with a total computing cost that is much smaller than that of the full log-likelihood in standard MCMC. They show that the asymptotic error of the resulting log-likelihood estimate is negligible even for a very small number of random samples $m$ ($m \ll n$), and demonstrate that (i) sub-sampling MCMC is substantially more efficient than standard MCMC in terms of sampling efficiency; and (ii) their approach outperforms other subsampling methods for MCMC proposed in the literature, including those listed at the beginning of this section.

Sub-sampling has important implications for MCMC inference. For example, in the standard MH sampler we accept a proposal draw with probability $u \sim Uniform(0,1)$ if and only if
\begin{equation}
\alpha = \frac{p(\theta^{\star} \vert y)q(\widehat{\theta}^{(i-1}) \vert \widehat{\theta}^{\star})}{p(\widehat{\theta}^{(i-1)} \vert y)q(\widehat{\theta}^{\star} \vert \widehat{\theta}^{(i-1)})} > u, \label{MH_evaluation}
\end{equation}
where we remind $\widehat{\theta}^{(i-1)}$ is the draw we have accepted in the previous iteration, and $\widehat{\theta}^{\star}$ the candidate draw in the current iteration, which will be accepted with probability $\alpha$. Evaluating repeatedly (in a Monte Carlo fashion) the expression in equation \eqref{MH_evaluation} using high-dimensional posterior densities, is quite cumbersome. By rearranging terms in this equation, taking logarithms, and splitting the likelihood function over the $N$ observations in the data $y$ we have
\begin{eqnarray}
\log \left \lbrace \frac{p(y \vert \widehat{\theta}^{\star} )}{p( y \vert \widehat{\theta}^{(i-1)}} \right \rbrace & > & \log \left \lbrace u \frac{q(\widehat{\theta}^{\star} \vert \widehat{\theta}^{(i-1)})}{q(\widehat{\theta}^{(i-1}) \vert \widehat{\theta}^{\star})} \right \rbrace  \Rightarrow \\
\frac{1}{N} \sum_{n=1}^{N} \log \left \lbrace \frac{p(y_n \vert \widehat{\theta}^{\star} )}{p( y_n \vert \widehat{\theta}^{(i-1)}} \right \rbrace & > &  \frac{1}{N} \log \left \lbrace u \frac{q(\widehat{\theta}^{\star} \vert \widehat{\theta}^{(i-1)})}{q(\widehat{\theta}^{(i-1}) \vert \widehat{\theta}^{\star})} \right \rbrace \Rightarrow \\
\frac{1}{N} \sum_{n=1}^{N} \lambda_{n}(\widehat{\theta}^{\star},\widehat{\theta}^{(i-1)}) & >  & c(u,\widehat{\theta}^{\star},\widehat{\theta}^{(i-1)}). \label{MH_subsampling}
\end{eqnarray}
Therefore, instead of sampling the full MH step in \eqref{MH_evaluation}, one can subsample the log-likelihood ratio quantity $\lambda_{n}(\widehat{\theta}^{\star},\widehat{\theta}^{(i-1)})$ and subsequently perform the approximate test $\lambda_{n}^{\star}(\widehat{\theta}^{\star},\widehat{\theta}^{(i-1)}) >  c(u,\theta^{\star},\widehat{\theta}^{(i-1)})$ in order to decide whether to accept $\widehat{\theta}^{\star}$ or not.

\subsection{Hamiltonian Monte Carlo}
Hamiltonian Monte Carlo (HMC) methods offer an alternative solution to the limitation of Metropolis-Hastings algorithm in exploring efficient high-dimensional posterior distributions. In particular, by carefully exploiting the differential structure of the target probability density, HMC provides an automatic procedure that yields a more efficient exploration of the probability space in such high dimensions. More specifically, HMC uses an approximate Hamiltonian dynamics simulation based on numerical integration which is then corrected by performing a Metropolis acceptance step.

In order to sample from the $K$-dimensional posterior distribution $p(\theta \vert y)$,  HMC introduces an independent $K$-dimensional auxiliary variable $\delta$ with density $p(\delta \vert \theta)$, which leads to the joint density 
\begin{align}
p(\theta,\delta) = p(\delta \vert \theta)p(\theta)
\end{align}
In most applications, including Stan, $p(\delta \vert \theta)$ is specified to be independent from the parameter vector $\theta$, for example using a multivariate normal distribution, i.e. $\delta \sim N(0,M)$, which leads to 
\begin{align}
p(\theta,\delta) = p(\theta) N(0,M)
\end{align} 
Let $H(\theta,\delta)$ denote the \textit{Hamiltonian} function, i.e. the negative joint log-probability, $H(\theta,\delta)= - \log p(\theta,\delta)$, and similarly let $\mathcal{L}(\theta)$ denote the logarithm of the target density $p(\theta)$. It can be shown (see Girolami and Calderhead, 2011) that 
\begin{align}
	H(\theta,\delta) = - \mathcal{L}(\theta) + \frac{1}{2}log\left\{(2\pi)^K\vert M\vert\right\} + \frac{1}{2} \delta^\prime M^{-1}\delta
\end{align} 
 
In practice, given a candidate draw $\delta^{(i)}$ from the $N(0,M)$ auxiliary density and the current draw $\theta^{(i)}$, the derivatives of $H(\theta,\delta)$ with respect to $\theta$ and $\delta$,
\begin{align}
\begin{split}
\frac{\partial H}{\partial \theta} &= -\mathcal{L}^\prime(\theta) \\
\frac{\partial H}{\partial \delta} &= M^{-1}\delta
\label{H_evol}
\end{split}
\end{align}
give rise to the transition $\theta^{(i)}  \rightarrow \theta^\ast$ and $\delta^{(i)}  \rightarrow \delta^\ast$. Next, the proposed $\theta^\ast$ (and $\delta^\ast$) are retained with probability
\begin{equation}
min \left\lbrace 1, \exp \left( H\left( \theta^{(i)}, \delta^{(i)}\right) - H\left( \theta^\ast, \delta^\ast\right) \right) \right\rbrace
\end{equation}
If the proposal is not accepted, the previous parameter value is returned for the next draw and used to initialize the next iteration.

\subsection{Parallelizing MCMC}
MCMC methods are characterized by the Markov property, that is, the fact that we need to first assess the current sample $\widehat{\theta}^{(i)}$ in order to decide whether $\widehat{\theta}^{(i+1)}$ is a possible sample from the target posterior. Therefore, due to this sequential dependence between iterations, it seems an oxymoron to attempt to parallelize \emph{across} MCMC iterations. As a consequence, a natural first step toward parallelization -- assuming we have a high-dimensional parameter $\theta$ that can be split into $r$ independent blocks $\theta_{r}$, $r=1,...,R$ -- would be to parallelize \emph{within} each iteration. That way we can compute each $p(\theta_{r} \vert y)$ in a separate worker. Malewicz et al. (2010) demonstrate such an algorithm in what is known as Google Pregel. However, Scott et al. (2016) note that not only such algorithms have very bad convergence rates, they are also extremely inefficient once one factors in computing costs and the marginal reductions in computing times.\footnote{For the Pregel environment in particular, a ten-fold increase in computing capacity only reduces computation time by a factor of two.} Similarly, Gonzalez et al. (2011) propose two parallel versions of the Gibbs sampler with good convergence guarantees, namely the Chromatic sampler and the Splash sampler. However, such parallel samplers are limited by the fact that there must be frequent (i.e. at each MCMC iteration) communication between the workers.

Instead of breaking a high-dimensional vector of parameters $\theta$ into smaller sub-vectors, Scott et al. (2016) propose to break the data $y$ into $R$ smaller blocks that can be distributed to an equivalent number of workers. This means that the high-dimensional posterior can be written as
\begin{equation}
p(\theta \vert y) = \prod_{r=1}^{R} p(\theta \vert y_r) \propto \prod_{r=1}^{R} p(y_{r} \vert \theta) p(\theta)^{1/R},
\end{equation}
where the prior is broken into $R$ independent components, $p(\theta) = \prod_{R} p(\theta)^{1/R}$ such that the total amount of prior information in the system is not affected by our decision to break $y$ in $R$ blocks.\footnote{One critique of this approach is that the prior may not provide enough regularization for each separate computation.} Assuming for simplicity that all workers each produce $S$ draws of $\theta$, then the consensus posterior will comprise $S$ draws that are weighted combinations of the $R \times S$ draws from all workers. Angelino et al. (2016, Section 4.2.1) provide citations to further studies that implement similar ideas towards the design of parallel MCMC.

An alternative way to exploit the idea of partitioning the data into $R$ non-overlapping subsets $y_{r}$, $r=1,...,R$, is to use the Weierstrass transform. For a function $f(\theta)$ the Weierstrass transform is a convolution of a Gaussian density with standard deviation $h$ and $f(\theta)$, and is of the form
\begin{equation}
W_h f(\theta) = \int_{- \infty}^{+ \infty} \frac{1}{\sqrt{2 \pi} h }  \exp \left \lbrace  - \frac{(\theta - \mu )^2}{2h^2}  \right \rbrace  f(\mu) d\mu. \label{Weirstrass_transform}
\end{equation}
$W_h f(\theta)$ can be thought of as a smooth approximation to $f(\theta)$.\footnote{When $h \rightarrow 0$, the transformation $W_h f(\theta)$ converges to $f(\theta)$.} Applying this transform to the posterior density, we get
\begin{eqnarray}
p(\theta \vert y) & = & \prod_{r=1}^{R} p(\theta \vert y_r) \approx \prod_{r=1}^{R} W_h p(\theta \vert y_r) \\
& = & \prod_{r=1}^{R}  \int_{- \infty}^{+ \infty} \frac{1}{\sqrt{2 \pi} h }  \exp \left \lbrace  - \frac{(\theta - \mu_{r} )^2}{2h^2} \right \rbrace p(\mu_{r} \vert y_r) d\mu_{r} \\
& \propto & \int_{- \infty}^{+ \infty} \prod_{r=1}^{R} \exp \left \lbrace  - \frac{(\theta - \mu_{r} )^2}{2h^2} \right \rbrace p(\mu_{r} \vert y_r) d\mu_{r}.
\end{eqnarray}
This last expression shows that after applying the Weirstrass transform, the posterior of $\theta$ can be viewed as the outcome of marginalizing latent parameters $\mu_{1},....,\mu_{R}$ from an augmented posterior $p(\theta,\mu_{1},....,\mu_{R} \vert y)$. This enables a subset-based Gibbs sampling, that is highly parallelizable, where we can first sample $\theta \vert \mu_{r},y \sim N(\widehat{\mu},h^2)$ and then $\mu_{r} \vert \theta,y \sim \frac{1}{\sqrt{2 \pi} h }  \exp \left \lbrace  - \frac{(\theta - \mu_{r} )^2}{2h^2}  \right \rbrace  p(\mu_{r} \vert y_r) $, see Wang and Dunson (2013) for more details on this scheme.

Other avenues of parallelizing MCMC do exist and their success depends on the inference problem at hand. For example, in problems with nonlinear coefficients whose posterior does not have a closed form expression, the Griddy Gibbs sampler of Ritter and Tanner (1992) can be used in order to evaluate such parameters in a grid (instead of sampling from their highly complex conditional posterior). The approximation in the Griddy-Gibbs sampler can be trivially parallelized, although the full algorithm itself can become very inefficient in high-dimensional models. Other examples include the Adaptive Griddy-Gibbs (AGG) and the Multiple-Try Metropolis (MTM) of Liu et al. (2000). Another related issue in MCMC methods is that of whether one needs to run a very long chain with as many iterations as (computationally) possible, or follow the advice of Gelman and Rubin (1992) and run several chains in parallel. Assuming random starting points, running chains in parallel allows to assess and speed up convergence by combining their output. Of course, when such chains run in parallel but are independent, the gains in efficiency are low. The Interchain Adaptive MCMC algorithm (Craiu et al., 2009) allows for parallel chains to interact within an adaptive Metropolis setting, such that substantial speed up in convergence is achieved. The adaptive element in this algorithm relies on the fact that each chain learns from its past, but also from the past iterations of other chains. Using this algorithm, Solonen et al. (2012) quote dramatic speed up in convergence by using only 10 chains in parallel. The ``Affine-Invariant'' Ensemble MCMC sampler of Goodman and Weare (2010) also involves parallel processing of chains in batches with efficiency gains in high dimensions. However, such samplers processing non-independent chains in parallel are restricted by the fact that communication between workers in a cluster must be frequent. Therefore, such samplers are slower than respective single-core MCMC samplers \emph{per iteration}, and computational gains from processing parallel chains only come from the fact that total convergence is achieved using a lower number of iterations.

\subsection{Maximum a posteriori estimation and the EM algorithm} \label{MAP}
Despite the increased availability of methods for making MCMC faster, there are cases where sampling from the full posterior might not be feasible or desired. As long as parameter uncertainty is not important for a specific empirical problem, one can work with point estimates that summarize some important features of the posterior distribution. In order to proceed with a point estimate $\widehat{\theta}$ of the unknown parameters $\theta$, we can introduce a cost function $\mathcal{C}$ that we aim to minimize. Therefore, the Bayesian equivalent of classical point estimation takes the following form
\begin{equation}
\arg \min_{\theta^\ast} \int \mathcal{C} (\theta - \theta^\ast) p(\theta \vert y) d\theta. \label{cost_function}
\end{equation}
It is trivial to show that when using the quadratic cost function $\mathcal{C} (\theta - \theta^\ast) = \left( \theta - \theta^\ast \right)^2$, equation \eqref{cost_function} is minimized for $\widehat{\theta} = \int \theta p(\theta \vert y) d\theta$, which is the posterior mean and is also known as the \emph{minimum mean square error (MMSE)} estimator. Similarly, the absolute cost function $\mathcal{C} (\theta - \theta^\ast) = \vert \theta - \theta^\ast \vert$ leads to the posterior median as the optimal point estimate.

An alternative point estimate can be obtained using the hit-and-miss cost function of the form
\begin{equation}
\mathcal{C} (\theta - \theta^\ast) = \left \lbrace  \begin{array}{l}
1, \text{ \ \ if }  \vert \theta - \theta^\ast \vert \geq \delta \\
0, \text{ \ \ if }  \vert \theta - \theta^\ast \vert < \delta
\end{array} \right.
\end{equation} 
for $\delta$ very small. Inserting this cost function in equation \eqref{cost_function} we obtain the solution
\begin{equation}
\widehat{\theta} = \arg \max_{\theta^\ast} p(\theta \vert y), \label{MAP_estimate}
\end{equation}
which is the posterior mode, also known as the \emph{maximum a posteriori (MAP)} estimator. Given (from Bayes rule) that $p(\theta \vert y) = p (y \vert \theta) p(\theta)$, it becomes apparent that Maximum Likelihood inference is a special case of MAP estimation with the uniform prior $p(\theta) \propto 1$.

MAP methods have been used in Bayesian inference for several decades. In a seminal paper, Tipping (2001) derives a MAP estimate of the parameters of a support vector machine model under a ``sparse Bayesian learning (SBL)'' prior. This prior for a parameter $\theta$ is a special case of a hierarchical Bayes structure where $\theta$ depends one some unknown hyperparameters $\xi$ that are random variables and have their own prior. Such hierarchical priors are used extensively nowadays in Bayesian analysis as a means of imposing shrinkage and computation is typically tackled by means of the Gibbs sampler (see Korobilis, 2013, for more details). In high-dimensional settings, however, sampling is not always feasible and Tipping (2001) derives a MAP estimator for the SBL prior using type-II maximum likelihood methods.\footnote{The name ``type-II maximum likelihood'' is a bit deceiving, as this method finds the value of the hyperparameter $\xi$ that maximizes the data marginal likelihood and not the likelihood function; see Berger (1985).}

Of course there are numerous ways one can solve the convex optimization problem in equation \eqref{MAP_estimate}, and we can't review all of them in such a short review. For example, Green et al. (2015) review proximal algorithms for obtaining the MAP estimate in high-dimensional settings; see also Parikh and Boyd (2013). Nevertheless, among all possible algorithms here we distinguish the EM algorithm. One reason for doing so is because the EM algorithm can be thought of as the optimization equivalent of the Gibbs sampler. Another important reason is that the EM algorithm is a unifying inference tool that nest several other approximating algorithms, such variational Bayes and message passing algorithms. There are several examples of high-dimensional MAP inference using the EM algorithm, and most notably we mention Rockova and George (2014).

\subsection{Variational Bayes and Expectation Propagation}
\subsubsection{Variational Bayes}
As in MAP inference, the main idea behind variational Bayes is to use optimization instead of sampling. First we introduce a family of densities $q(\theta)$ and subsequently we try to find a certain density $q^{\star}(\theta)$ that minimizes the Kullback-Leibler (KL) divergence to the exact posterior $p(\theta \vert y)$. Mathematically we want to minimize the following function
\begin{eqnarray}
q^{\star}(\theta) &= & \arg \min_{q\left(\theta\right)} KL(q \Vert p) \\
& = &\arg \min_{q\left(\theta\right)} \int q(\theta) \log \left \lbrace \frac{q(\theta)}{p(\theta \vert y) } \right \rbrace d\theta,
\end{eqnarray}
where it holds that $KL(q \Vert p) \geq 0$, with value equal zero only when $q(\theta)$ is identical to the true posterior $p(\theta \vert y)$. It can be shown that this minimization problem is equivalent to finding a $q(\theta)$ that maximizes the marginal likelihood. This is because for the logarithm of the marginal likelihood holds
\begin{eqnarray}
\log \left( p(y ) \right) & = & \log \left( p(y ) \right) \int q(\theta) d\theta = \int q(\theta) \log( p( y )) d\theta \\
& = & \int q(\theta) \log \left \lbrace  \frac{p(y,\theta) / q(\theta)}{p(\theta \vert y) / q(\theta)} \right \rbrace d\theta \\
& = & KL + \int q(\theta) \log \left \lbrace \frac{p(y,\theta)}{ q(\theta)} \right \rbrace d\theta,
\end{eqnarray}
which, given that $KL \geq 0$, gives
\begin{equation}
p(y) \geq \exp \left( \int q(\theta) \log \left \lbrace \frac{p(y,\theta)}{ q(\theta)} \right \rbrace d\theta \right).
\end{equation}
Therefore, the VB optimization problem becomes that of maximizing the lower bound for the marginal likelihood. Note that this problem is different from MAP because here we are looking to optimize with respect to a function $q(\theta)$ and not just the random variable $\theta$. For that specific reason, this optimization problem for the functional $q(\bullet)$ can be solved iteratively using calculus of variations. Before we do so, it is convenient to split $\theta$ into $J$ independent blocks, i.e. $q(\theta) = \prod_{j=1}^{J} q(\theta_{j})$.\footnote{This decomposition is called the \emph{mean-field approximation}, a term originating from Physics.} Then we can show that $p(y)$ can be maximized by iterating sequentially through
\begin{eqnarray}
q^{\star}(\theta_{1}) & \propto & \exp \left( \int \log p(y,\theta) p(\theta_{(-1)}) d \theta_{(-1)} \right), \\
\vdots \nonumber \\
q^{\star}(\theta_{J}) & \propto & \exp \left(  \int \log p(y,\theta) p(\theta_{(-J)}) d \theta_{(-J)} \right),
\end{eqnarray}
where $\theta_{(-j)}$ denotes $\theta$ with its $j^{th}$ element removed. It turns out that this iterative scheme is very similar to the EM algorithm. Each integral provides the expectation of the joint posterior with respect to the density $p(\theta_{(-j)})$ for all $j=1,...,J$. Loosely speaking, this scheme also resembles a Gibbs sampler. However, instead of sampling, we fix $\theta_{(-j)}$ to their posterior mean values.
\subsubsection{Expectation Propagation}
Expectation propagation (EP) is related to variational Bayes, but it can be considered as a separate class of algorithms. In contrast to VB, EP attempts to minimize the ``reverse'' KL divergence measure
\begin{equation}
KL = \int p(\theta \vert y) \log \left \lbrace \frac{p(\theta \vert y)}{q(\theta)} \right \rbrace d\theta. \label{EP_problem}
\end{equation}
We showed previously that the variational Bayes optimization problem leads to calculating expectations with respect to the proposal density $q( \theta )$ (after we split $\theta$ into independent blocks). In contrast, the EP optimization problem shown in equation \eqref{EP_problem} can be thought of as requiring to take expectation with respect to the unknown posterior $p(\theta \vert y)$. For that reason the EP optimization approach is different to VB. 

First, we assume that the joint distribution can be decomposed into $N$ ``factors'' of the form
\begin{equation}
p ( \theta, y )  = \prod_{n=1}^{N} f_n(\theta).
\end{equation}
Next we need to choose $q(\bullet)$ based on the exponential family of distributions, and assume that this is also decomposed into $N$ factors of the form
\begin{equation}
q(\theta) = \frac{1}{Z} \prod_{n=1}^{N} \widetilde{f}_n(\theta),
\end{equation}
where $Z$ is a normalizing constant that makes the distribution integrate to one. The idea is to process the EP optimization problem for each of the $N$ factors separately.\footnote{By splitting the problem into $N$ factors/batches, it should become apparent that EP algorithms can be trivially parallelized.} Each factor $\widetilde{f}_n(\theta)$ is refined \emph{iteratively} by making $q^{\star}(\theta) \propto \widetilde{f}_n(\theta) \prod_{j=1,j \neq n}^{N} \widetilde{f}_j(\theta) $ a closer approximation to $p^{\star}(\theta) \propto f_n(\theta) \prod_{j=1,j \neq n}^{N} \widetilde{f}_j(\theta)$. Then $\widetilde{f}_n(\theta)$ is removed from the approximating distribution by calculating $q^{\star\star}(\theta) = q(\theta) / \widetilde{f}_n(\theta)$ and we define $p^{\star\star} (\theta) = \frac{1}{Z^{\star}} f_n(\theta)q^{\star\star}(\theta)$. In the final step of the iterative scheme, the factor $\widetilde{f}_n(\theta)$ is updated such that the sufficient statistics of $q(\theta)$ match those of $p^{\star\star} (\theta)$.

While the description provided is very generic, implementations of expectation propagation can take several interesting forms depending on the application. The loopy belief propagation algorithm that is used to compute marginal posterior distributions in Bayesian networks is a special case of EP, as are other cases of the general class of \emph{message passing algorithms}.\footnote{In computing science message passing is the concept of depicting graphically, typically using graphical models, how the parameters and the factors (functionals) interact with each other. The resulting class of algorithms can be extremely powerful and trivially parallelizable; see Korobilis (2020) for more details.} Such algorithms are at the forefront of statistical and machine learning research in the Big Data era.

\subsection{Approximate Bayesian Computation}
In high-dimensional applications, with high complexity and volume of available data, calculation of the likelihood or the posterior might be computationally intractable or closed-form expressions might not be available. There are also cases in fields such as image analysis or epidemiology where the normalizing constant of the likelihood is unknown. Approximate Bayesian Computation (ABC) is specifically appropriate for use in such cases. Therefore, the argument in favor of ABC is not only that it is more computationally efficient than MCMC methods, rather it can be used in many complex problem when application of MCMC is infeasible. 

A basic version of ABC, that provides $n$ samples of the parameter of interest $\theta$, can be summarized with the following pseudo-algorithm \\ \\
\noindent \textbf{\underline{Basic ABC rejection sampler}}
\begin{itemize}
\item[ ] for $i = 1:n$
\begin{itemize}
\item[ ] repeat
\begin{itemize} 
 \item Generate a $\widehat{\theta}^{\star}$ randomly from the prior $p(\theta)$
\item Generate randomly data $z$ using the specified econometric model, with $\theta$ fixed to the generated value $\widehat{\theta}^{\star}$
\end{itemize}
\item[ ] until $\rho \left(z,y \right) \leq \epsilon$
\item[ ] set $\widehat{\theta}^{(i)} = \widehat{\theta}^{\star}$

\end{itemize}
\item[ ] end for
\end{itemize}
In this algorithm, $\rho \left(z,y \right)$ is a distance function (e.g. Euclidean) measuring how close the generated data $z$ are relative to the observed data $y$, and $\epsilon \rightarrow 0$. In the case of high-dimensional data, the probability of generating a $z$ that is close to $y$ goes to zero. Therefore, in practice ABC algorithms evaluate the distance between summary statistics of $z$ and $y$. In this case we would evaluate instead the distance function $\rho \left( \eta\left(z \right),\eta\left( y \right)\right) \leq \epsilon$, where $\eta(\bullet)$ is a function defining a statistic which most often is not sufficient. Using summary statistics may result in loss of accuracy, especially in cases where not many summary statistics of a dataset are available.

The above scheme samples $\theta$ from the approximate posterior
\begin{eqnarray}
p_{\epsilon} \left( \theta \vert y \right) & = & \int p_{\epsilon} \left( \theta,z \vert y \right) dz \\
&= & \int \frac{p(\theta) \times p \left( z \vert \theta \right) \mathscr{I} \left(\rho \left( \eta\left(z \right),\eta\left( y \right)\right) \leq \epsilon \right)}{ p(z) \mathscr{I} \left(\rho \left( \eta\left(z \right),\eta\left( y \right)\right) \leq \epsilon \right)} dz \\
& \approx & \frac{p(\theta) p(y \vert \theta)}{p(y)} \equiv p(\theta \vert y),
\end{eqnarray}
where $\mathscr{I} \left( A \right)$ is a function that takes the value one if expression $A$ holds, and it is zero otherwise. 

An obvious problem with this scheme is that it heavily relies on the choice of prior. Particularly in high-dimensional settings, using simulated values from the prior $p(\theta)$ is inefficient and results in proposals that are located in low probability regions of the true posterior we want to approximate. In this case we can define the following MCMC-ABC algorithm, which is a likelihood free MCMC sampler \\ \\
\noindent \textbf{\underline{MCMC-ABC algorithm}}
\begin{itemize}
\item[ ] for $i = 1:n$
\begin{itemize}
\item[ ] repeat
\begin{itemize} 
 \item Generate $\widehat{\theta}^{\star}$ from a proposal distribution $q\left(\theta \vert \theta^{(i-1)} \right)$
\item Generate $z$ from the likelihood $p \left(y \vert \widehat{\theta}^{\star} \right)$
\item Generate $u$ from $\mathcal{U}_{[0,1]}$ and compute the acceptance probability $$\alpha \left( \widehat{\theta}^{\star},\widehat{\theta}^{(i-1)} \right) = min \left\lbrace 1, \frac{p(\widehat{\theta}^{\star})q\left(\widehat{\theta}^{(i-1)} \vert \widehat{\theta}^{\star}  \right)}{p(\widehat{\theta}^{(i-1)})q\left(\widehat{\theta}^{\star} \vert \widehat{\theta}^{(i-1)}  \right)} \right \rbrace$$ 
\end{itemize}
\item[ ] if 
\begin{itemize}
\item[ ] $u \leq \alpha \left( \widehat{\theta}^{\star},\widehat{\theta}^{(i-1)} \right)$ and $\rho \left(z,y \right) \leq \epsilon$, set $\widehat{\theta}^{(i)} = \widehat{\theta}^{\star}$
\end{itemize}
\item[ ] else 
\begin{itemize}
\item[ ] set $\widehat{\theta}^{(i)} = \widehat{\theta}^{(i-1)}$
\end{itemize}
\item[ ] end if
\end{itemize}
\item[ ] end for
\end{itemize}
The algorithm is not literally speaking ``likelihood-free'' as the likelihood is used in order to generate $z$. However, the likelihood is not used in order to calculate the acceptance probability $\alpha \left( \widehat{\theta}^{\star},\widehat{\theta}^{(i-1)} \right)$.

ABC can be extended in several interesting ways, for example combined with sequential Monte Carlo, or they can incorporate model selection in a trivial way.\footnote{The posterior probability of a given model can be approximated by the proportion of accepted simulations given the model.} As with variational Bayes, ABC has experienced immense growth in mainstream statistics over the past two decades, and our prediction is that it will also soon be embraced by economists in order to solve complex problems.\footnote{See for example Frazier, Maneesoonthorn, Martin and McCabe (2019) for an application of ABC algorithms in producing financial forecasts in computationally efficient ways.}

\newpage
\section{Non-algorithmic ways of speeding up Bayesian inference}
The purpose of this Section is to build further intuition by demonstrating various ways to approximate a high-dimensional inference problem simply by re-writing the likelihood and facilitating computation.\footnote{Another factor that affects computation is the choice of programming language and the way one interacts with it. However, discussing such details is beyond the scope of our review.} There are specific problems where just by simply re-writing the likelihood in an equivalent form we can gain a lot in computation -- especially when Bayesian sampling methods are used to approximate the posterior (such as traditional MCMC methods). Of course there are numerous examples of such approaches in the literature, and we only selectively quote some tools we have favored ourselves while trying to develop new estimation algorithms. We provide a few examples from some popular classes of models in economics, namely regressions with many predictors and large vector autoregressions.

\subsection{Random projection methods}

Random projection methods have been used in fields such as machine learning and image recognition as a way of projecting the information in data sets with a huge number of variables into a much lower dimensional set of variables. To fix the basic ideas of random projections, let $X$ be a $T\times k$ data matrix involving $T$ observations on $k$ variables where $k\gg T$. $X_{t}$ is a $1\times k$ vector denoting the $t^{th}$ row of $X$. Define the projection matrix, $\Phi ,$ which is $m\times k$ with $m\ll k$ and $
\widetilde{X}_{t}^{\prime }=\Phi X_{t}^{\prime }$. Then $\widetilde{X}_{t}$
is the $1\times m$ vector denoting the $t^{th}$ row of the compressed data
matrix, $\widetilde{X}$. Since $\widetilde{X}$ has $m$ columns and $X$ has $%
k $, the former is much smaller and is much easier to work with. To see how
this works in a regression context, let $y_{t}$ be a scalar dependent variable
and consider the relationship: 
\begin{equation}
y_{t}=X_{t}\beta +\varepsilon _{t}.  \label{regression}
\end{equation}%
If $k\gg T$, then working directly with (\ref{regression}) is impossible
with some statistical methods (e.g. maximum likelihood estimation) and
computationally demanding with others (e.g. Bayesian approaches which require the use of MCMC methods). Some of the computational burden can arise simply due to the need to store in memory huge data matrices. For instance, calculation of the Bayesian posterior mean under a natural conjugate prior requires, among other manipulations, inversion of a $k\times k$ matrix involving the data. This can be difficult if $k$ is huge. In order to deal with a large number of predictors, one can specify a compressed regression variant of (\ref{regression}) 
\begin{equation}
y_{t}=\widetilde{X}_{t}\beta ^{c}+\varepsilon
_{t}.  \label{compressed_regression}
\end{equation}%
Once the explanatory variables have been compressed (i.e. conditional on $\Phi $), standard Bayesian regression methods can be used for the regression of $y_{t}$ on $\widetilde{X}_{t}$. If a natural conjugate prior is used, then analytical formulae exist for the posterior, marginal likelihood, and predictive density, and computation is trivial. 

\medskip 

Note that the model in \eqref{compressed_regression} has the same structure as a reduced-rank regression, as the $k$ explanatory variables in the original regression model are squeezed into a small number of explanatory variables given by the vector $\widetilde{X}_{t}^{\prime}=\Phi X_{t}^{\prime }$. The crucial assumption is that $\Phi $ is not estimated from the data, rather it is treated as a random matrix with its elements sampled using random number generation schemes.\footnote{For that reason, random projection methods are referred to as \textit{data oblivious}, since $\Phi $ is drawn without reference to the data.} The underlying motivation for random compression arises from the Johnson-Lindenstrauss lemma. This states that any $k$ point subset of the Euclidean space can be embedded in\ $m=O\left( \log \left( k\right) /\epsilon ^{2}\right) $ dimensions without distorting the distances between any pair of points by more than a factor of $1\pm \epsilon $, where $0<\epsilon <1$. There are various ways to draw $\Phi $; most obviously we can generate this matrix from N(0,1) or a Uniform(0,1) distributions. Alternatively we can draw $\Phi _{ij}$, the $ij^{th}$ element of $\Phi $, (where $i=1,..,m$ and $j=1,..,k$) from the following scheme that generates a sparse random projection 
\begin{equation}
\begin{tabular}{l}
$\Pr \left( \Phi _{ij}=\frac{1}{\sqrt{\varphi }}\right) =\varphi ^{2}$ \\ 
$\Pr \left( \Phi _{ij}=0\right) =2\left( 1-\varphi \right) \varphi $ \\ 
$\Pr \left( \Phi _{ij}=-\frac{1}{\sqrt{\varphi }}\right) =\left( 1-\varphi
\right) ^{2}$%
\end{tabular}%
,  \label{PHI}
\end{equation}%
where $\varphi $ and $m$ are unknown parameters.\footnote{The Johnson-Lindenstrauss lemma suggests that $\Phi $ should be a random matrix whose columns have unit lengths and, hence, Gram-Schmidt orthonormalization is done on the rows of the matrix $\Phi $.} While the remarkable properties of random compression hold even for a single, data oblivious, random draw of $\Phi $, in practical situations (e.g. forecasting) we would like to ensure that we work with random projections that are optimal in a data-rigorous sense. As long as each compressed model projected with the matrix $\Phi$ can be estimated very quickly (e.g. using natural conjugate priors), then one should be able to generate many random projections and estimate simultaneously many small models. Then goodness-of-fit measures can be used to assess which compressed models (corresponding to different random projections) fits the data better.

\medskip 

In summary, huge dimensional data matrices (that are too large to insert to standard econometric models) can be compressed quickly into a much lower dimension by generating random projections, without the cost of solving some computationally expensive optimization problem. The resulting compressed data matrix can then be used in a statistical model such as a regression or a vector autoregression, that can be estimated easily with traditional estimation tools. This very general approach has excellent potential applications in numerous problems in economics. For an application in large vector autoregressions and for further references, see Koop et al. (2019).

\subsection{Variable elimination in regression}
Variable elimination or marginalization is a machine learning procedure used in graphical models that, loosely speaking, allows (via certain rules) to break a high-dimensional inference problem into a series of smaller problems. We can use similar ideas in a standard regression setting in order to facilitate high-dimensional inference. Assume that we work again with a regression model setting with $p$ predictors, but this time interest lies in the $j$-th predictor and its coefficient. We can rewrite the regression as
\begin{equation}
y = x_{j}\beta_{j} + x_{\left(-j\right)} \beta_{\left(-j\right)} + \varepsilon, \label{new_regression}
\end{equation}
where $y$, $x_{j}$ and $\varepsilon$ are all $T \times 1$ vectors and $x_{\left(-j\right)}$ is a $T \times (p-1)$ predictor matrix with predictor $j$ removed. It might be the case that we are interested only in parameter $\beta_{j}$ because this is a policy parameter. A first useful result is the one of partitioned regression: defining the $T\times T$ annihilator matrix $M_j = I_T - x_j\left(x_j^\prime x_j\right)^{-1} x_j^\prime$, it is easy to show using the algebra of partitioned matrices that $\widehat{\beta}_j$, the OLS estimates of $\beta_j$ can be obtained as the solution of
\begin{align}
	\widehat{\beta}_j = \left(x_j^\prime x_j\right)^{-1} x_j^\prime\left(y-x_{(-j)}\widehat{\boldsymbol{\beta}}_{(-j)}\right)
	\label{part_reg_1}
\end{align}
where the sub-vector $\widehat{\beta}_{(-j)}$ is the solution of the following regression 
\begin{align}
	\widehat{\beta}_{(-j)} = \left(x_{(-j)}^{\dagger \prime}x_{(-j)}^{\dagger} \right)^{-1}x_{(-j)}^{\dagger \prime} y^{\dagger}
	\label{part_reg_2}
\end{align}
with $x_{(-j)}^{\dagger}=M_j x_{(-j)}$ and $y^{\dagger}=M_j y$ denoting the projections of $x_{(-j)}$ and $y$ on a space that is orthogonal to $x_j$.

This result provides very useful intuition about the relationships between our variables and coefficients in the OLS regression. Most importantly they can be generalized to efficient procedures for high-dimensional inference. Consider for example combining partitioned regression results with a penalized estimator instead of OLS. To demonstrate this point, we consider an alternative partition of the regression. Define the $T \times 1$ vector $q_{j} = x_{j}/\Vert x_{j}\Vert$, and generate randomly a matrix $Q_j$ that is normalized as $Q_{j} Q_{j}^{\prime} = I - q_j q_j^\prime$. This means that the matrix $Q = [q_j,Q_j]$ is orthogonal, such that multiplying both sides of \eqref{new_regression} by $Q^{\prime}$ gives
\begin{eqnarray}
Q^{\prime}y & = &  Q^{\prime}x_{j}\beta_{j} + Q^{\prime}x_{\left(-j\right)} \beta_{\left(-j\right)} + Q^{\prime}\varepsilon \Rightarrow \\
\left[ \begin{array}{c}
q_{j}^{\prime}y \\
Q_{j}^{\prime}y
\end{array}\right] & = & \left[ \begin{array}{c}
q_{j}^{\prime}x_{j} \\
Q_{j}^{\prime}x_{j}
\end{array}\right] \beta_j + \left[ \begin{array}{c}
q_{j}^{\prime}x_{(-j)} \\
Q_{j}^{\prime}x_{(-j)}
\end{array}\right] \beta_{(-j)} + Q^{\prime} \varepsilon \Rightarrow   \\
\left[ \begin{array}{c}
y^{\ast} \\
y^{+}
\end{array}\right] & = & \left[ \begin{array}{c}
\Vert x_{j} \Vert \\
 0
\end{array}\right] \beta_j + \left[ \begin{array}{c}
x_{(-j)}^{\ast} \\
x_{(-j)}^{+}
\end{array}\right] \beta_{(-j)} + \widetilde{\varepsilon}, \label{rotated}
\end{eqnarray}
where $y^{\ast}=q_{j}^{\prime}y$, $y^{+}=Q_{j}^{\prime}y$, $x_{(-j)}^{\ast}= q_{j}^{\prime}x_{(-j)}$, $x_{(-j)}^{+} = Q_{j}^{\prime}x_{(-j)}$ and $\widetilde{\varepsilon} = Q^{\prime} \varepsilon $. In this derivation we have used the fact that $Q_{j}^{\prime}x_{j}= Q_{j}^{\prime} q_{j}\Vert x_j \Vert = 0$ because $Q_j$ and $q_j$ are orthogonal. Additionally, $var(\widetilde{\varepsilon}) = \sigma^{2}Q^{\prime}Q=\sigma^{2}=var(\varepsilon)$ because by construction $Q^{\prime}Q=I$. The likelihood of the transformed regression model in equation \eqref{rotated} is multivariate Normal, which means we can use standard results for conditional Normal distributions to show that we can first estimate $\beta_{(-j)}, \sigma^{2}$ by regressing $y^{+}$ to $x_{(-j)}^{+}$, and then at a second stage obtain $\beta_{j}$ by regressing $y^{\ast}$ on $\Vert x_{j} \Vert$ conditional on $\beta_{(-j)}, \sigma^{2}$ being known. This is a very useful result since now, conditional on obtaining in a first step some estimates of $\beta_{(-j)}, \sigma^{2}$, we can estimate $\beta_{j}$ in a regression with known variance.\footnote{Most importantly, we can do so in parallel for all predictors $j=1,...,p$.} Korobilis and Pettenuzzo (2019) apply these ideas to a high-dimensional VARs under a wider class of hierarchical shrinkage priors. Considering that the exact way of calculating marginal posteriors would involve solving numerically a $p-1$-dimensional integral for each $j$, doing a rotation of the form shown above, and deriving the marginal posteriors analytically, means large gains in computation can be achieved.

\subsection{Multivariate estimation equation-by-equation}
Some of the most important quantitative exercises that policy-makers are interested in, involve the vector autoregressive (VAR) model and its variants. Economic theories can be tested reliably only in a multivariate econometric setting, and the same holds to a large degree for measuring the impact of shocks to the wider economy. While a large part of empirical analysis is done using VARs of say three or five variables, there is an expanding literature that acknowledges the benefits of large VARs. In particular, small structural VARs might not be invertible meaning that their residuals will not span the same space as the structural shocks that macroeconomists want to identify. Therefore, it comes to no surprise that there is an expanding and lively literature on methods for estimating large VARs.

A vector autoregression for an $1 \times n$ vector of variables of interest $y_{t}$ can be written in the following form
\begin{equation}
y_{t} = B_{0} + \sum_{i=1}^{p} y_{t-i}B_{i} + \varepsilon_{t},
\end{equation}
but we can write it in familiar multivariate regression form as
\begin{equation}
y_{t} = X_{t} B + \varepsilon_{t},  \label{VAR_reg}   
\end{equation}
where $X_{t} = \left(1,y_{t-1},...,y_{t-p}\right)$, $A = \left[B_0,B_1,...,B_p\right]$ and $\varepsilon_{t} \sim  N\left(0, \Sigma \right)$ with $\Sigma$ and $n \times n$ covariance matrix. Accumulation of parameters in VARs is quite different compared to univariate models. A VAR with $n=3$ variables, intercept terms and $p=1$ lag has 18 parameters. The same VAR with $n=50$ variables has 3825 parameters. The last VAR with $p=12$ has 31,325 parameters. This gives an idea of the polynomial rate at which the number of parameters increases as $n$ and/or $p$ increase. The problem with VARs proliferates if we want to use independent priors on the coefficients $B_{i}$ that would allow to shrink each of their elements independently. Doing so implies that we need to write the VAR in seemingly unrelated regression (SUR) form, where in this form the right hand side matrix of predictors is $Z = I_n \otimes X$. For large VARs this $T \times n(np+1)$ matrix becomes so large that handling it eventually becomes computationally infeasible, despite the fact that it is sparse and one can rely on more efficient sparse matrix calculations. 

Nevertheless, there are still simple ways to use independent priors. Koop et al. (2019) in the context of developing random projection algorithms for large VARs, proposed to break the VAR into a collection of $n$ univariate equations. Using ideas from estimation of simultaneous equation models we can transform the VAR in triangular form. Consider the Cholesky-like decomposition of the covariance matrix, $\Sigma = A^{-1}D \left(A^{-1}\right)^{\prime}$ where $D$ is a diagonal matrix for variances, and $A^{-1}$ is a uni-triangular matrix of the form
\begin{equation}
\boldsymbol{A}^{-1}=\left[ 
\begin{array}{ccccc}
1 & 0 & ... & 0 & 0 \\ 
\alpha _{2,1} & 1 & \ddots & \vdots & \vdots \\ 
\vdots & \ddots & \ddots & 0 & 0 \\ 
\alpha _{n-1,1} & ... & \alpha _{n-1,n-2} & 1 & 0 \\ 
\alpha _{n,1} & ... & \alpha _{n,n-2} & \alpha_{n,n-1} & 1%
\end{array}%
\right].
\end{equation}%
Under this decomposition we can rewrite the VAR in equation \eqref{VAR_reg} as
\begin{eqnarray}
y_{t} & = & X_{t} B + u_{t}\left(A^{-1}D^{\frac{1}{2}}\right)^{\prime}  \Rightarrow \\
y_{t}A & = & X_{t} BA + u_{t},D^{\frac{1}{2}}  \Rightarrow \\
y_{t} + y_{t}\widetilde{A}& = & X_{t} \Gamma + u_{t},D^{\frac{1}{2}} \Rightarrow \\ 
y_{t} & = & X_{t}\Gamma - y_{t}\widetilde{A} + u_{t},D^{\frac{1}{2}}, \label{triang1}
\end{eqnarray}
where $u_{t} \sim N(0,I)$, $\Gamma = B\times A$ and $\widetilde{A} = A - I$ is a lower diagonal matrix created from $A$ after we remove its unit diagonal elements. This is a so-called triangular VAR system due to the fact that $\widetilde{A}$ has a lower triangular structure. It cannot be estimated as a multivariate regression using standard linear estimators because $y_{t}$ shows up both on the left-hand side and the right-hand side of the equation. However, due to the lower triangular structure of $\widetilde{A}$ and the fact that $D$ is diagonal the system can be estimated equation-by-equation using simple OLS. This means that in high dimensions we can essentially write the VAR in this form and apply any univariate regression estimator and algorithm we like.\footnote{Of course, note that this flexibility comes at the cost of shrinkage or variable selection being dependent on the ordering of the variables in the VAR; see Koop et al. (2019) for a discussion.} More importantly, note that the last equation shows that all contemporaneous covariances among the $n$ VAR equations can be written as RHS predictors $-y_{t}$. This is an important implication because it shows that $\widetilde{A}$ can be treated as a regression parameter and (given that we can estimate these equations recursively) we can readily apply methods of the previous section to impose shrinkage also on the VAR covariance matrix.

Finally, we can derive a similar triangular VAR that has slightly different representation and implications for estimation. Begin with equation \eqref{VAR_reg} but now rewrite it in the form
\begin{eqnarray}
y_{t} & = & X_{t} B + u_{t}\left(A^{-1}D^{\frac{1}{2}}\right)^{\prime}  \Rightarrow \\
y_{t} & = & X_{t} B + u_{t}\left(\left(\widetilde{A}^{-1} + I\right) D^{\frac{1}{2}}\right)^{\prime}   \Rightarrow \\
y_{t} & = & X_{t} B + u_{t}\widetilde{A}^{-1}D^{\frac{1}{2}} + u_{t}D^{\frac{1}{2}} \Rightarrow \\ 
y_{t} & = & X_{t} B + v_{t}\widetilde{A}^{-1} + v_{t}, \label{triang2}
\end{eqnarray}
where $v_{t} \sim N(0,D)$ and $\widetilde{A}^{-1} = A^{-1} - I$ is a triangular matrix created by removing the identity diagonal of $A^{-1}$. This system can also be estimated equation by equation, where in equation $i$ we use residuals from the previous $i-1$ equations. This form has different implications for designing estimation algorithms compared to the one in \eqref{triang1}, even though they are observationally equivalent. Equation \eqref{triang2} allows direct estimation of the VAR matrices $B$ and $A^{-1}$, while equation \eqref{triang1} estimates functions of those, i.e. $\Gamma$ and $A$. Such examples show that high-dimensional inference can be approximated by efficient transformations of the VAR model that allow to readily apply univariate estimators which are simpler and possibly algorithmically faster.

\section{Conclusions}
We have attempted to provide a wide review of algorithms and methods for speeding Bayesian inference to cope with high-dimensional data and models. Our review is very high-level and should be seen as a first-step introduction to the various tools that a modern econometricians need to have in their toolbox. As always, there are several pros and cons with the various algorithms, and the choice of the ``\textit{right'}' algorithm is application specific. There are some excellent and in-depth recent reviews of some of these algorithms that demonstrate their use in various interesting contexts. For example, Angelino et al. (2016) and Green et al. (2015) provide some excellent detailed reviews of various algorithms. Blei et al. (2010) provide an accessible introduction to variational Bayes methods. Sisson et al. (2018) provide a recent review and references of Approximate Bayesian Computation (ABC) methods. The review paper by Zhu et al. (2017) focuses on scalability and distributed computation of Monte Carlo methods, as well as regularized Bayesian inference. Bayesian machine learning is a very lively literature, as is the case with non-Bayesian machine learning approaches that are also expanding rapidly. We have tried to provide a gentle introduction to this literature and bridge the gap between the expanding computing needs of economists and computational advances proposed in various other literatures such as comprehensive sensing, computer vision and AI.

\newpage
\singlespacing

\end{document}